# In-Context Learning for Label-Efficient Cancer Image Classification in Oncology

Mobina Shrestha[1], Bishwas Mandal[2], Vishal Mandal[3], Asis Shrestha[4]

Sentara Albemarle Medical Center, USACS Elizabeth City, NC[1]
Kansas State University[2]
CSX Corp[3]
UAMS Winthrop P. Rockefeller Cancer Institute[4]

**ABSTRACT**

The application of AI in oncology has been limited by its reliance on large, annotated datasets and the need for retraining models for domain-specific diagnostic tasks. Taking heed of these limitations, we investigated in-context learning as a pragmatic alternative to model retraining by allowing models to adapt to new diagnostic tasks using only a few labeled examples at inference, without the need for retraining. Using four vision-language models (VLMs)—Paligemma, CLIP, ALIGN and GPT-4o, we evaluated the performance across three oncology datasets: MHIST, PatchCamelyon and HAM10000. To the best of our knowledge, this is the first study to compare the performance of multiple VLMs on different oncology classification tasks. Without any parameter updates, all models showed significant gains with few-shot prompting, with GPT-4o reaching an F1 score of 0.81 in binary classification and 0.60 in multi-class classification settings. While these results remain below the ceiling of fully fine-tuned systems, they highlight the potential of ICL to approximate task-specific behavior using only a handful of examples, reflecting how clinicians often reason from prior cases. Notably, open-source models like Paligemma and CLIP demonstrated competitive gains despite their smaller size, suggesting feasibility for deployment in computing constrained clinical environments. Overall, these findings highlight the potential of ICL as a practical solution in oncology, particularly for rare cancers and resource-limited contexts where fine-tuning is infeasible and annotated data is difficult to obtain.

**INTRODUCTION**

Artificial intelligence (AI) is beginning to play a meaningful role in diagnostic medicine, with growing interest in its potential to support clinical decision-making and improve disease classification [1,2]. The success of AI models in medical imaging, particularly in radiology and pathology has demonstrated that AI systems can achieve expert-level performance in tasks like tumor detection, histopathology-based subtyping, and biomarker prediction [3–6]. However, these achievements have largely relied on supervised learning techniques that require large amounts of annotated datasets for each specific task. This reliance on annotated data and task-specific retraining remains a major bottleneck, particularly for rare diseases and resource-constrained settings where such datasets are scarce or difficult to generate [7–9].

An alternative to this process is in-context learning (ICL), a paradigm originally introduced in the field of natural language processing (NLP). Instead of retraining models for every new task, ICL enables foundation models to generalize using a few labeled examples provided at inference time [10]. This few-shot approach parallels clinical reasoning to some extent, where clinicians reference prior cases to guide diagnostic decisions without requiring formal retraining. Recent work has demonstrated the utility of ICL in medical NLP tasks, including clinical question answering, differential diagnosis generation, and evidence summarization [11,12]. However, many of these ICL-based systems have remained limited to text-based tasks. This text-based, or more broadly unimodal focus overlooks the inherently multimodal nature of clinical practice, where diagnostic reasoning often relies on the integration of diverse data types such as imaging, histopathology, and clinical notes [13–15]. The implementation of vision-language models (VLMs) can help bridge this gap to some extent. VLMs such as GPT-4o, Paligemma and LLaVA are capable of processing and reasoning across both visual and textual contexts, offering the potential to apply in-context learning to multimodal clinical contexts [16–19]. Similarly, there have been some studies that demonstrate that these models can generalize across a range of diagnostic tasks using only a small number of examples, without retraining [25, 20, 21]. Such an approach could be transformative in clinical contexts where data availability is limited [9,22,23]. Also, few-shot multimodal learning may offer a solution to the persistent challenges of data imbalance and underrepresentation in medical AI research [24]. In this study, we evaluated the performance of four vision-language models viz. GPT-4o, Paligemma, CLIP, and ALIGN across three cancer datasets: MHIST (colorectal polyps) [26], PatchCamelyon (breast cancer metastasis) [27-28], and HAM10000 (skin lesions) [29]. We experimented with both zero-shot and few-shot settings, assessing whether these models could generalize across various cancer subtypes using minimal examples. Overall, our findings suggest that few-shot prompting, even without fine-tuning, can enable foundation models to achieve diagnostic performance comparable to traditional supervised methods. The results suggest that, in some clinical contexts, the use of generalist foundation models may reduce the need for developing separate, task-specific models, particularly in oncology, where limited data and disease heterogeneity poses significant challenges.

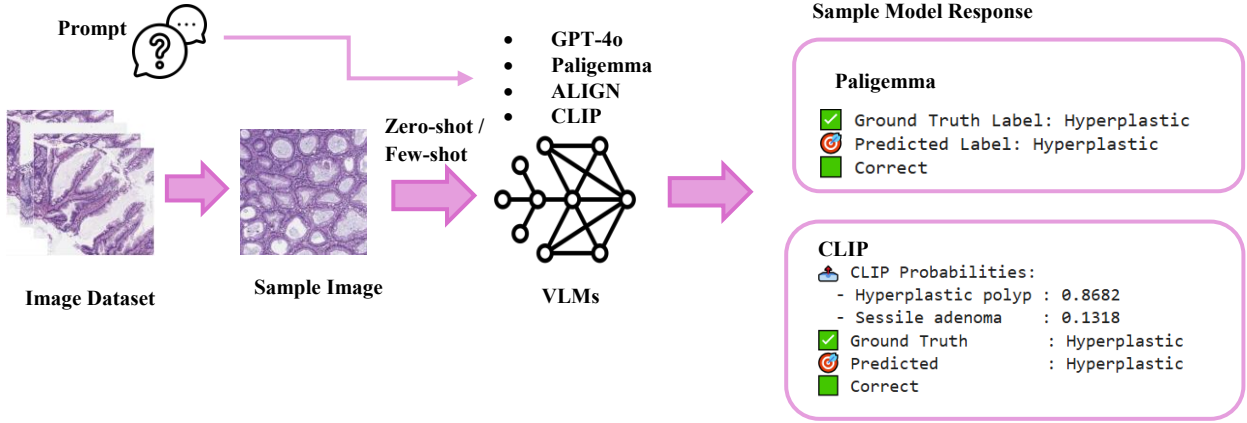

Figure 1. Schematic overview of the zero-shot and few-shot classification of cancer images using VLMs.

## METHODS

### Study Design and Data Sources

This study was conducted using publicly available data and does not include confidential information. All research procedures were conducted in accordance with the Declaration of Helsinki, asserting required ethical standards. The diagnostic performance of four VLMs, i.e. GPT-4o, Paligemma, CLIP, and ALIGN were evaluated across three oncology datasets as shown in Table 1. All models were prompted to select the correct diagnosis from a predefined multiple-choice list for each test image. The classification tasks included three clinical scenarios: MHIST required distinguishing between hyperplastic polyps and sessile serrated adenomas in colorectal samples; PatchCamelyon (PCam) involved identifying the presence or absence of tumor in breast lymph node histology; and HAM10000 required models to classify the correct skin lesion out of seven possible choices, as shown in HAM10000 classes row in Table 1.

| Dataset | MHIST | PatchCamelyon | HAM10000 |
| --- | --- | --- | --- |
| Description | Histopathology images of colorectal polyps stained with hematoxylin and eosin | Histopathological scans of lymph node sections. Breast cancer metastases vs. normal lymph nodes | Dermatoscopic images of common pigmented skin lesions |
| Classes | Hyperplastic Polyp (HP); Sessile Serrated Adenoma (SSA) | Tumor; No Tumor | melanoma; melanocytic nevus; basal cell carcinoma; actinic keratosis; benign keratosis; dermatofibroma and vascular lesion |
| Image Resolution | 224 x 224 pixels | 96 x 96 pixels | 450 x 600 pixels |
| Dataset Size | 3,152 | 327,680 | 10,015 |
| Evaluation Set | 158 | 158 | 161 |

Table 1. Overview of the three histopathology benchmarking datasets.

For MHIST and PatchCamelyon, the evaluation set consisted of 158 images per dataset, with 79 images per class. For HAM10000, the evaluation set contained 161 images, with 23 images across seven classes.

**Model Implementation and Inference Setup**
Experiments for GPT-4o were conducted via the OpenAI Python API [32]. For Paligemma, we used the pretrained checkpoint (google/paligemma-3b-mix-224) [30] from the Hugging Face model hub, configured for vision-language inference. CLIP was evaluated using the ViT-B/32 variant (openai/clip-vit-base-patch32) [31] and an open-source implementation consistent with the ALIGN architecture (kakaobrain/align-base) [33] was used to approximate ALIGN's functionality. For simplicity, we refer to this model as "ALIGN" throughout this study. The same implementation of ALIGN was previously applied by Mandal et al [36]. During experimentation, we explored different values for temperature and top_p hyperparameters, with values ranging from 0.5 to 0.7 to evaluate their impact on prediction consistency and reasoning quality. All models were run in their native vision-language configurations, capable of processing paired image and text inputs. No additional fine-tuning or parameter updates were performed for any of the models. Similarly, balanced evaluation sets were generated using random sampling to maintain equal class representation across all experimental conditions. In few-shot settings, 3, 5, and 10 support examples per class were randomly selected and embedded within the prompt.

**Systematic Bias Mitigation and Prompt Engineering**
To mitigate potential biases and ensure robust evaluation, support examples were randomly sampled from the training data by excluding the test set images. We implemented option randomization which shuffled label positions across test instances to counter primacy and recency effects [34]. For binary classification tasks (MHIST and PCam), the two options were dynamically reordered per inference query, while the seven HAM10000 labels underwent Latin square permutation to maintain balanced position distributions. This approach aligns with survey methodology adaptations for AI systems, reducing habitual model preferences for specific options [34]. We made sure that all evaluations used standard multiple-choice question formats, with clear instructions to use "select ONLY one diagnosis from the following options" prompt to constrain output variability. Likewise, we maintained our prompt engineering strategy by drawing inspiration from evidence-based techniques relevant to clinical AI prompting [35]. We applied chain-of-thought prompting strategies following all known best practices and prompting tricks (e.g., "Analyze stepwise: (1) Assess glandular architecture, (2) Evaluate cytological features…") to mirror diagnostic reasoning patterns observed in medical AI frameworks. For policy-constrained models such as GPT-4o and ALIGN, prompts were framed as clinical consultations ("As a pathologist reviewing this case…") to bypass alignment filters while maintaining task relevance.

**Performance Evaluation and Statistical Analysis**
Model performance was evaluated using the weighted F1 score as the primary metric. For binary classification tasks on MHIST and PatchCamelyon datasets, the precision, recall, and F1 score were reported with reference to the positive class, i.e. Sessile adenoma for MHIST and Tumor for PatchCamelyon. For HAM10000, which involved seven classes, a weighted F1 score was calculated. To assess the reliability of the model estimates, 95% confidence intervals (CIs) were generated for the weighted F1 scores using bootstrap resampling with 10,000 iterations. In each iteration, the test dataset was resampled with replacement, and the weighted F1 score was recalculated.

**RESULTS**
**Few-shot prompting improves model performance across histopathology and dermatology tasks**
We evaluated the performance of few-shot prompting of four generalist VLMs across three cancer datasets. A high-level overview of datasets used in the study is described in Table 1. Performance was assessed in both zero-shot and few-shot (3, 5, and 10-shot) settings. In MHIST, weighted F1 scores in the zero-shot setting ranged from 0.57 for Paligemma (95% CI: 0.51–0.62) and CLIP (CI: 0.50–0.65) to 0.60 for GPT-4o (CI: 0.54–0.65) and 0.59 for ALIGN (CI: 0.52–0.65). With 3-shot prompting, all models improved: GPT-4o reached 0.66 (CI: 0.58–0.75), Paligemma 0.66 (CI: 0.59–0.73), CLIP 0.63 (CI: 0.54–0.69), and ALIGN 0.62 (CI: 0.57–0.69). Further gains were seen at 5-shot and 10-shot. At 10-shot, GPT-4o achieved a weighted F1 score of 0.81 (CI: 0.75–0.85), Paligemma 0.79 (CI: 0.70–0.83), ALIGN 0.72 (CI: 0.66–0.77), and CLIP 0.70 (CI: 0.65–0.75). In the PatchCamelyon dataset, a similar pattern was observed. Zero-shot weighted F1 scores ranged from 0.52 for CLIP (CI: 0.46–0.55) and ALIGN (CI: 0.42–0.57) to 0.56 for Paligemma (CI: 0.46–0.59) and 0.59 for GPT-4o (CI: 0.48–0.65). At 3-shot, GPT-4o improved to 0.64 (CI: 0.54–0.69), Paligemma to 0.63 (CI: 0.51–0.67), CLIP to 0.59 (CI: 0.48–0.65), and ALIGN to 0.56 (CI: 0.50–0.60). With 10-shot prompting, GPT-4o reached 0.79 (CI: 0.70–0.85), followed by Paligemma 0.75 (CI: 0.66–0.81), ALIGN 0.77 (CI: 0.65–0.82), and CLIP 0.72 (CI: 0.63–0.79). In the HAM10000 dataset, baseline zero-shot performance was lower across all models, reflecting the challenge of multi-class classification.

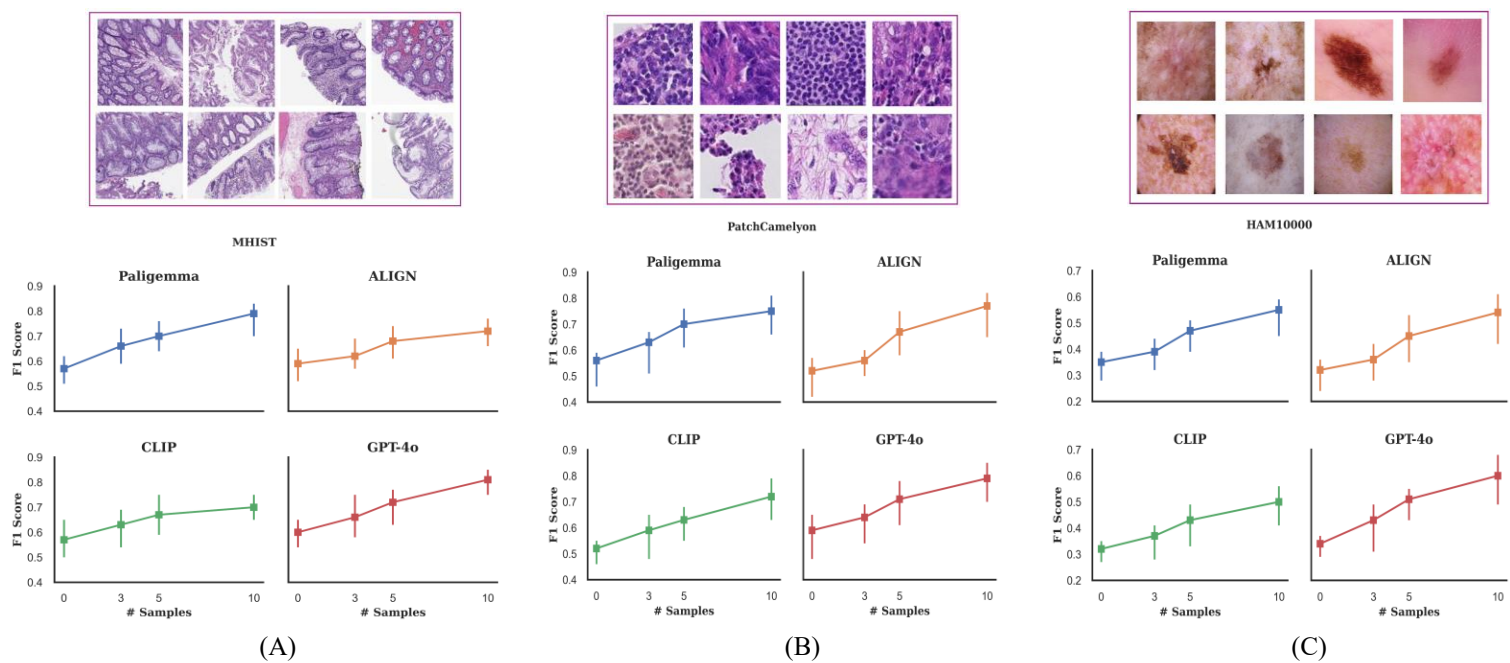

**Figure 2.** Zero and few-shot learning performance of four vision-language models (Paligemma, ALIGN, CLIP, GPT-4o) across three cancer image datasets: (A) MHIST, (B) PatchCamelyon, and (C) HAM10000. GPT-4o consistently achieved the highest F1 scores as the number of samples increased.

Weighted F1 scores at zero-shot were 0.34 for GPT-4o (CI: 0.29–0.37), 0.35 for Paligemma (CI: 0.28–0.39), 0.32 for CLIP (CI: 0.27–0.35), and 0.32 for ALIGN (CI: 0.24–0.36). Few-shot prompting resulted in consistent improvements. With 3-shot prompting, GPT-4o improved to 0.43 (CI: 0.31–0.49), Paligemma to 0.39 (CI: 0.32–0.44), CLIP to 0.37 (CI: 0.28–0.41), and ALIGN to 0.36 (CI: 0.28–0.42). By 10-shot, GPT-4o reached 0.60 (CI: 0.49–0.68), Paligemma 0.55 (CI: 0.45–0.59), ALIGN 0.54 (CI: 0.42–0.61), and CLIP 0.50 (CI: 0.41–0.56). Across all datasets, GPT-4o consistently achieved the highest absolute performance, though Paligemma, CLIP, and ALIGN demonstrated parallel improvements. The relative gains with few-shot prompting were particularly notable in the HAM10000 dataset, where multi-class complexity initially limited zero-shot performance. These results highlight that few-shot in-context learning substantially enhances the diagnostic performance of generalist VLMs across binary and multi-class medical imaging tasks. A complete summary of performance metrics, including F1 scores, precision, recall, and confidence intervals, is provided in Table 2.

**Model performance in binary classification tasks (MHIST and PCam)**
In MHIST dataset, hyperplastic polyps were classified with moderate accuracy in the zero-shot setting. GPT-4o achieved the highest F1 score (0.69), followed closely by Paligemma (0.68), ALIGN (0.67), and CLIP (0.65). Recall values for hyperplastic polyps ranged between 0.60–0.75 across models. Sessile serrated adenomas, however, were more difficult to identify, with zero-shot F1 scores ranging between 0.45 to 0.52. Even in regular clinical practice, differentiating SSAs from HPs on hematoxylin and eosin-stained samples is a challenge given the subtle and overlapping histomorphology features. However, few-shot prompting demonstrated substantial performance improvements for both SSAs and HPs. After 10-shot prompting, GPT-4o achieved an F1 score of 0.83 for HPs and 0.79 for SSAs. Paligemma and ALIGN similarly improved. In the PatchCamelyon dataset, which involved distinguishing tumor from non-tumor lymphatic tissue, zero-shot F1 scores were moderate, ranging from 0.51–0.61 for tumor detection across models. Paligemma and GPT-4o showed higher baseline recall (0.61–0.64), whereas ALIGN and CLIP exhibited lower initial sensitivity. With increasing few-shot samples, tumor detection performance improved substantially where GPT-4o reached an F1 score of 0.78 and Paligemma attained 0.77 after 10-shot prompting. However, classification of non-tumor samples remained more variable. While GPT-4o achieved a non-tumor F1 score of 0.80 at 10-shot, Paligemma and ALIGN models sometimes favored higher sensitivity at the expense of specificity, misclassifying reactive lymphoid tissue as tumor. Particularly at lower shot numbers, Paligemma's F1 score for non-tumor was 0.54 in the zero-shot setting before improving with prompting.

| Dataset | Model | Zero-shot | | | 3-shot | | | 5-shot | | | 10-shot | | |
|---|---|---|---|---|---|---|---|---|---|---|---|---|---|
| | | F-1 score | Precision | Recall | F-1 score | Precision | Recall | F-1 score | Precision | Recall | F-1 score | Precision | Recall |
| MHIST | Paligemma | 0.57 | 0.57 | 0.59 | 0.66 | 0.60 | 0.72 | 0.70 | 0.71 | 0.71 | 0.79 | 0.73 | 0.85 |
| | 95% CI | (0.51-0.62) | | | (0.59-0.73) | | | (0.64-0.76) | | | (0.70-0.83) | | |
| | ALIGN | 0.59 | 0.53 | 0.66 | 0.62 | 0.62 | 0.64 | 0.68 | 0.77 | 0.62 | 0.72 | 0.73 | 0.72 |
| | 95% CI | (0.52-0.65) | | | (0.57-0.69) | | | (0.61-0.74) | | | (0.66-0.77) | | |
| | CLIP | 0.57 | 0.56 | 0.61 | 0.63 | 0.65 | 0.63 | 0.67 | 0.70 | 0.66 | 0.70 | 0.71 | 0.71 |
| | 95% CI | (0.50-0.65) | | | (0.54-0.69) | | | (0.59-0.75) | | | (0.65-0.75) | | |
| | GPT-4o | 0.60 | 0.61 | 0.62 | 0.66 | 0.67 | 0.67 | 0.72 | 0.68 | 0.77 | 0.81 | 0.81 | 0.82 |
| | 95% CI | (0.54-0.65) | | | (0.58-0.75) | | | (0.63-0.77) | | | (0.75-0.85) | | |
| PatchCamelyon | Paligemma | 0.56 | 0.53 | 0.59 | 0.63 | 0.64 | 0.63 | 0.70 | 0.69 | 0.73 | 0.75 | 0.74 | 0.77 |
| | 95% CI | (0.46-0.59) | | | (0.51-0.67) | | | (0.61-0.76) | | | (0.66-0.81) | | |
| | ALIGN | 0.52 | 0.58 | 0.48 | 0.56 | 0.50 | 0.64 | 0.67 | 0.68 | 0.68 | 0.77 | 0.77 | 0.79 |
| | 95% CI | (0.42-0.57) | | | (0.50-0.60) | | | (0.58-0.75) | | | (0.65-0.82) | | |
| | CLIP | 0.52 | 0.58 | 0.48 | 0.59 | 0.65 | 0.54 | 0.63 | 0.60 | 0.68 | 0.72 | 0.67 | 0.78 |
| | 95% CI | (0.46-0.55) | | | (0.48-65) | | | (0.55-0.68) | | | (0.63-0.79) | | |
| | GPT-4o | 0.59 | 0.62 | 0.57 | 0.64 | 0.64 | 0.66 | 0.71 | 0.66 | 0.76 | 0.79 | 0.79 | 0.80 |
| | 95% CI | (0.48-0.65) | | | (0.54-0.69) | | | (0.61-0.78) | | | (0.70-0.85) | | |
| HAM10000 | Paligemma | 0.35 | 0.35 | 0.37 | 0.39 | 0.38 | 0.42 | 0.47 | 0.47 | 0.47 | 0.55 | 0.58 | 0.53 |
| | 95% CI | (0.28-0.39) | | | (0.32-0.44) | | | (0.39-0.51) | | | (0.45-0.59) | | |
| | ALIGN | 0.32 | 0.34 | 0.31 | 0.36 | 0.38 | 0.36 | 0.45 | 0.47 | 0.46 | 0.54 | 0.52 | 0.57 |
| | 95% CI | (0.24-0.36) | | | (28-0.42) | | | (0.35-0.53) | | | (0.42-0.61) | | |
| | CLIP | 0.32 | 0.32 | 0.32 | 0.37 | 0.35 | 0.41 | 0.43 | 0.42 | 0.46 | 0.50 | 0.52 | 0.50 |
| | 95% CI | (0.27-0.35) | | | (0.28-0.41) | | | (0.33-0.49) | | | (0.41-0.56) | | |
| | GPT-4o | 0.34 | 0.37 | 0.33 | 0.43 | 0.42 | 0.45 | 0.51 | 0.48 | 0.55 | 0.60 | 0.57 | 0.64 |
| | 95% CI | (0.29-0.37) | | | (0.31-0.49) | | | (0.43-0.55) | | | (0.49-0.68) | | |

**Table 2.** Metrics for VLMs on zero-shot and few-shot settings across three cancer datasets

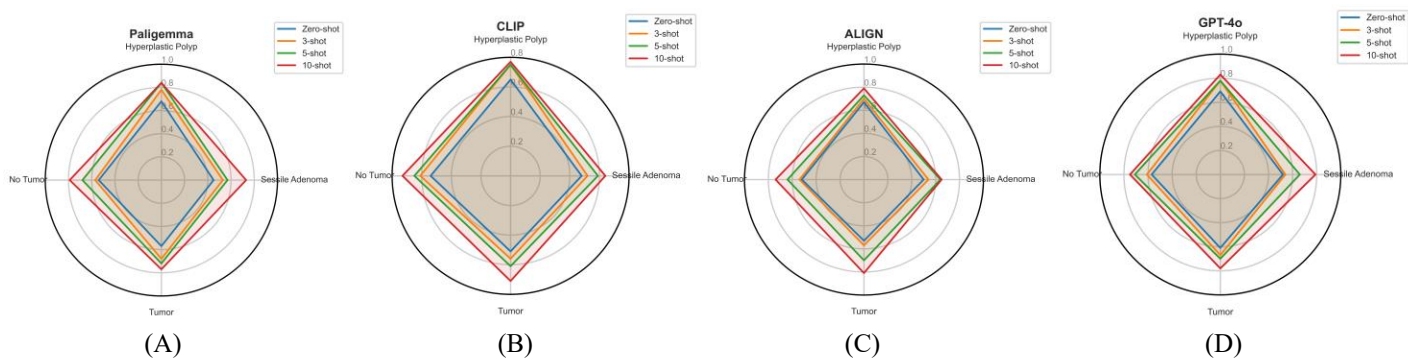

(A)      (B)      (C)      (D)

**Figure 3.** Class-wise F1 scores for (A) Paligemma, (B) CLIP, (C) ALIGN, and (D) GPT-4o across MHIST dataset: Hyperplastic Polyp, Sessile Adenoma and PatchCamelyon: Tumor, and No Tumor. Each radar plot displays performance across zero-shot, 3-shot, 5-shot, and 10-shot learning settings.

**Model performance in multi-class classification tasks (HAM10000)**

Multi-class classification of skin lesions in the HAM10000 dataset remains a challenging task, largely due to the morphological overlap and visual similarities among the seven lesion types. This complexity was reflected in the zero-shot performance, with relatively lower F1 scores, ranging from 0.26 to 0.38 across all lesion categories. With few-shot, substantial gains were observed across all models.

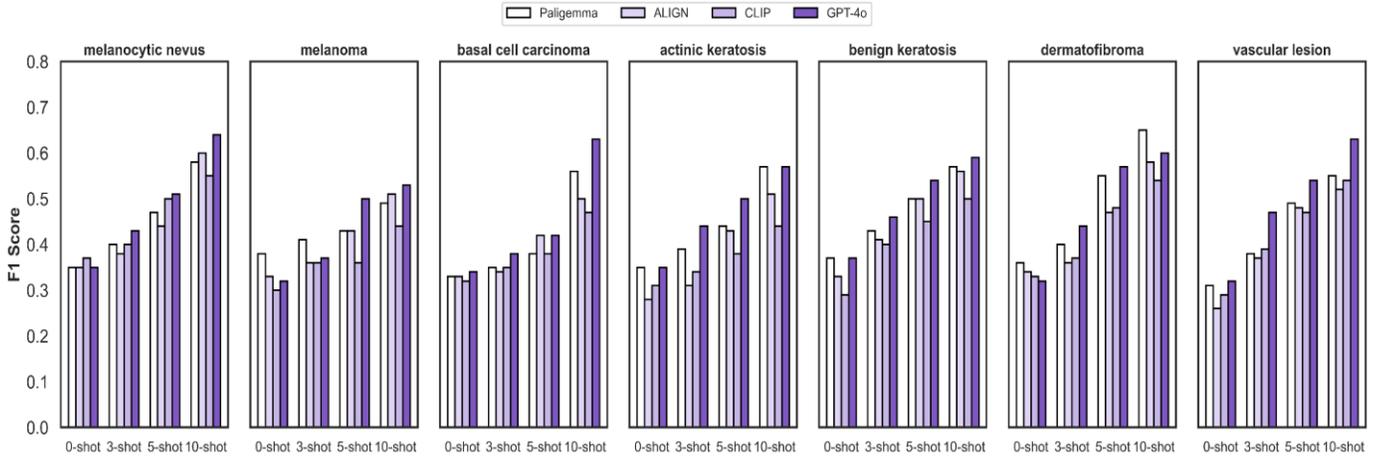

**Figure 4**. Comparison of F1 scores across zero-shot and few-shot settings for multi-class skin lesion classification using four VLMs (Paligemma, ALIGN, CLIP, and GPT-4o) on the HAM10000 dataset. Each panel represents a distinct lesion type. Overall, GPT-4o consistently outperformed other models across all shot settings and lesion categories, with performance improving as the number of shots increases.

For melanocytic nevi, Paligemma demonstrated an increase in F1 score from 0.35 at zero-shot to 0.58 at 10-shot while GPT-4o achieved a slightly superior performance with the F1 score rising from 0.35 to 0.64. In melanoma classification, GPT-4o improved its F1 score from 0.32 (zero-shot) to 0.53 (10-shot), compared to Paligemma's increase from 0.38 to 0.49. GPT-4o also demonstrated superior recall in melanoma at 10-shot (0.58 versus 0.44 for Paligemma), implying a higher sensitivity critical for early malignant detection. For basal cell carcinoma, GPT-4o again outperformed Paligemma at the 10-shot level, achieving an F1 score of 0.63 compared to 0.56, with recall values of 0.67 and 0.61 respectively.

Similarly, in the classification of actinic keratosis, GPT-4o improved from an F1 score of 0.35 at zero-shot to 0.57 at 10-shot, with recall increasing from 0.30 to 0.64. Paligemma, while improving as well (F1 from 0.35 to 0.57), lagged slightly in recall (0.63 vs. 0.64 for GPT-4o). GPT-4o achieved an F1 score of 0.59 for benign keratosis and 0.60 for dermatofibroma, whereas Paligemma scored 0.51 and 0.60 respectively. Interestingly, for dermatofibroma, both models reached similar F1 levels, although GPT-4o exhibited higher recall (0.66 versus 0.64). Vascular lesions posed a greater challenge overall, with baseline zero-shot F1 scores around 0.26–0.32; however, GPT-4o achieved the highest F1 at 10-shot (0.63), accompanied by a recall of 0.68, outperforming Paligemma (F1 score 0.55, recall 0.52). ALIGN and CLIP models demonstrated similar trends of improvement but generally underperformed relative to GPT-4o and Paligemma across all lesion types. For example, CLIP's 10-shot F1 scores ranged from 0.44 to 0.54 across categories, while ALIGN's F1 scores ranged from 0.37 to 0.52, with particularly lower recall values in malignant lesions.

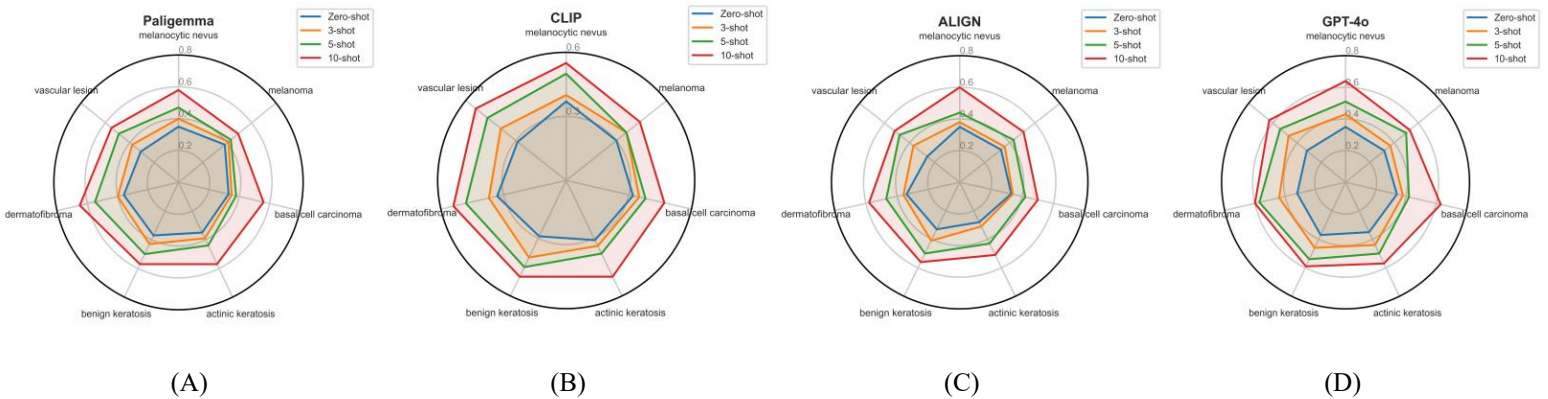

(A)          (B)          (C)          (D)

**Figure 5.** Comparison of VLMs: (A) Paligemma, (B) CLIP, (C) ALIGN, and (D) GPT-4o on the HAM10000 skin lesion classification task. Radar plots show how performance improves with more support examples across lesion categories.

Overall, the results demonstrated the difficulty of predicting a multi-class classification task while also highlighting the value of in-context learning for enhancing diagnostic accuracy in fine-grained and overlapping lesion categories.

**In-Context Learning for Sample-Efficient Adaptation in Oncology**
We implemented ICL by constructing prompts that paired n-support examples, i.e. n ∈ {0, 3, 5, 10} with their corresponding text labels, followed by an unlabeled query image. For models capable of processing image-text sequences (i.e. Paligemma and GPT-4o), each support example was represented as a concatenated visual and natural language descriptor. For image-text matching models (i.e. CLIP and ALIGN), we adapted the ICL structure to a retrieval-based formulation, aggregating similarity scores between the query image and class-conditioned prompts informed by the support set. Across all datasets and VLMs, ICL yielded consistent improvements in performance metrics. For instance, in HAM10000, Paligemma demonstrated a progression in weighted F1 score from 0.35 (zero-shot) to 0.55 (10-shot), while GPT-4o improved from 0.34 to 0.60 (CI:0.49-0.68) over the same range. These trends were reflected in MHIST, where Paligemma increased its weighted F1 score from 0.57 (zero-shot) to 0.79 (10-shot), and in PCam, where GPT-4o rose from 0.59 to 0.79 (CI:0.70-0.85). The performance gains were more pronounced in underrepresented classes such as dermatofibroma and vascular lesions in HAM10000, where ICL appeared to mitigate prior class imbalance by reinforcing visual priors relevant to low-frequency phenotypes.

The results demonstrate that ICL facilitates a form of localized task conditioning, wherein the model selectively re-weighs attention toward class-relevant features based on few-shot vision-language examples. This reflects an instance of amortized inference, where the model sidesteps task-specific gradient updates and instead modulates its internal representations by integrating support set derived signals with its pretrained knowledge. This mechanism parallels clinical diagnostic reasoning, which often relies on pattern recognition from previously encountered cases which is particularly crucial in low-prevalence or ambiguous scenarios. By instantiating this process through ICL, we enable large-scale pretrained models to generalize in a sample-efficient manner, minimizing the need for downstream fine-tuning or parameter updates. This is particularly useful in oncology, where annotated datasets can be scarce. By integrating few-shot examples into the model's decision-making process, ICL enables more efficient use of available samples, improving diagnostic accuracy in conditions where class imbalance or low-prevalence diseases are common.

**SUMMARY**
This study explores the application of ICL using VLMs for cancer image classification. While the achieved performance of approximately 0.8 F1 score for binary classification and around 0.6 for seven-class classification does not surpass that of fully fine-tuned models, the findings underscore the practical advantages of ICL in oncology. One of the primary benefits of ICL is its ability to function effectively with minimal labeled data. This is particularly relevant in oncology, where acquiring large amounts of annotated datasets is often challenging due to HIPAA concerns, the need for expert annotation, and the rarity of certain cancer types. By leveraging a few examples within prompts, ICL enables models to generalize and make accurate predictions without extensive retraining. This approach aligns with the findings of Ferber et al. [25], where they demonstrated that GPT-4V, when provided with a limited number of examples, could achieve performance comparable to specialized neural networks trained on specific tasks. Also, ICL offers flexibility by allowing models to be applied across various tasks without the need for task-specific fine-tuning. This is particularly advantageous in clinical settings where rapid deployment and adaptability are crucial. The ability to utilize generalist AI models trained on non-domain specific dataset democratizes access to advanced diagnostic tools, enabling medical professionals without software engineering or AI background to benefit from AI-driven insights.

Despite these advantages, some limitations of ICL remain evident. The observed performance ceiling, particularly in the multi-class setting, reflects the limitations of relying solely on few-shot examples to capture the diagnostic subtleties present in complex oncologic imaging. Many cancers subtypes exhibit overlapping morphologies that are difficult to ascertain without richer contextual information than what prompts alone can offer. Also, the effectiveness of ICL is closely tied to the design of the prompt itself, so support examples must be carefully selected and structured, as even small variations can lead to inconsistent outputs which were observed in our experiments. So, in a way this dependence introduces variability that is less prominent in fully supervised models. Computational demands also remain a barrier, although models like GPT-4o offer strong multimodal capabilities, their inference costs and hardware requirements may not be feasible for all clinical environments. In contrast, our results show that open-source, lighter-weight models such as Paligemma and CLIP still benefit substantially from ICL, providing viable, scalable options for institutions with limited computational resources. Overall, these findings suggest that while ICL may not fully displace fine-tuned models in every case, it represents a pragmatic and inclusive strategy for deploying AI in oncology particularly in low-resource settings, rare disease contexts, or applications requiring rapid task adaptation without retraining.